\documentclass[amsmath,amssymb,aps,nofootinbib]{revtex4}
\usepackage[dvips]{graphicx}
\usepackage{graphicx}
\usepackage{amsfonts}
\usepackage{bm}
\usepackage{amsmath}
\usepackage{amssymb}
\usepackage{color}
\usepackage[all]{xy}

\newcommand{\rd}{\mathrm{d}} 
\newcommand{\ri}{\mathrm{i}} 

\begin{document}

\title{Topological Characterization of Higher Dimensional Charged Taub--NUT Instantons}
\author{Daniel \surname{Flores-Alfonso}$^{1,}$}
\email[]{daniel.flores@correo.nucleares.unam.mx}
\author{Hernando \surname{Quevedo}$^{1,2,3}$}
\email[]{quevedo@nucleares.unam.mx}
\affiliation{$^1$Instituto de Ciencias Nucleares,
Universidad Nacional Aut\'onoma de M\'exico,\\
AP 70543, Ciudad de M\'exico 04510, M\'exico \\
$^2$Dipartimento di Fisica and ICRA, Universit\`a di Roma "La Sapienza", I-00185 Roma, Italy\\
$^3$Institute of Experimental and Theoretical Physics,
Al-Farabi Kazakh National University, 
Almaty 050040, Kazakhstan}

\begin{abstract}
Recently, we have shown that non-selfdual self-gravitating dyonic fields with magnetic mass generalize the Dirac monopole.
The unique topological index, which characterizes the field, is a four dimensional analogue of the famous monopole configuration.
An unexpected result of this analysis is that the electric parameter can only take certain discrete values as a consequence of applying the 
path integral approach to quantize the magnetic flux. Here, we show how this result can be generalized to higher dimensions,
considering a special type of inhomogeneous geometries. Our results apply to a vast range of theories and situations in which topological charges are present. For concreteness, we focus here on Lovelock--Maxwell solutions and show that the 
magnetic flux corresponds to a topological excitation and the electric flux becomes discrete.
\\[0.2cm]
\end{abstract}

\maketitle

\label{first}

\section{Introduction}

The idea of using topological tools to derive quantum properties of classical systems is not new and has been applied actively in the context of monopole and instanton configurations \cite{frankel}.  Initially thought as pure theoretical artifacts, topological charges 
are now widely accepted as corresponding, in most cases, to true physical quantities.  
One example are winding numbers, which are known to represent the quantum nature of many physical quantities and have been verified in several 
physical systems; for instance, in superfluids, the winding number of a condensate wave function is related to
vorticity flux quanta. Moreover, the circularity of superfluid Helium can only take integer
multiples of $h/m$, where $m$ is the mass of a $^4$He atom or a pair of $^3$He atoms.
The winding of the phase is a topological feature and since the flow momentum in superfluids
is given essentially by the differential of the phase, then circularity is quantized.
Topological structures of this type yield a quantization of the magnetic charge $p$ of an isolated monopole \cite{naber1,wuyang}.
The interaction of this bosonic monopole field with a fermion of electric charge $q_F$ leads to the condition 
that $q_F$ can only take discrete values \cite{dirac}. These values are integer multiples of $\hbar/2p$ and are a consequence of
demanding that the wave function is globally defined on the line bundle associated to the monopole's underlying U(1)-bundle. 
These examples show that the study of topological charges of different physical configurations can be of special importance
for the investigation and understanding of physical measurable quantities. More recently, a novel method has been proposed by which
topological charges, in general, can be detected by using quench dynamics \cite{zzl18}. 

Certain areas of quantum and semiclassical physics have been enriched by algebraic topology. Our goal in this paper is to
apply mathematical results of bundle theory to objects of Euclidean quantum gravity. In general, quantum gravity remains an outstanding problem of theoretical physics, although several methods and techniques 
have been proposed to analyze this problem \cite{carlip}. For instance, the path integral approach uses Wick rotated classical solutions
to derive the thermodynamics of black holes. To do so, one imposes regularity conditions on Euclidean metrics and performs 
a saddle point approximation. This method yields the Hawking temperature of black holes and coincides
with the calculations of semiclassical gravity, where quantum fields are considered over curved classical backgrounds.

In a previous work \cite{dfhq}, we examined Taub--NUT gravitational instantons with Wick rotated Maxwell fields.
The magnetic flux was found to be a topological excitation of the gauge field given essentially by a winding number.
The regularity conditions imposed to perform a saddle point approximation lead to the quantization of the electric charge.
In this work, we generalize these results. Inhomogeneous geometries of Einstein manifolds serve as way to define
higher dimensional Taub--NUT spaces. The metrics are defined over complex line bundles whose base space is a K\"ahler
manifold. The four dimensional Taub--NUT metric \cite{taub,nut,misner} belongs to this class of geometries. Although,
instead of a complex line a Lorentzian plane is the standard fiber of the bundle space, e.g., spacetime.

Our motivation for focusing on Taub--NUT metrics stems from two different scenarios which both relate the
metric to magnetic monopoles. Firstly, the asymptotic behavior of the magnetic Weyl tensor \cite{ellis} of the Lorentzian Taub--NUT geometry
corresponds to a monopolar source. The NUT parameter is analogous to magnetic charge and parallels the r\^ole played by mass in the 
electric Weyl tensor. Hence, the NUT parameter is a sort of magnetic mass,
which has no Newtonian analogue. Secondly, the selfdual Euclidean Taub--NUT metric serves as the spatial sector
of a five dimensional spacetime \cite{hawking} . The Kaluza--Klein procedure shows that a four dimensional observer perceives this
five dimensional spacetime as a four dimensional scenario with a Dirac monopole present \cite{sorkin}. In summary,
comparison with the Dirac configuration has yielded gravitomagnetic sources, 
Kaluza--Klein monopoles and topological charges in gravitational settings, each related to a distinct version
of the Taub--NUT metric. Higher dimensional Kaluza--Klein monopoles have been known to exist for some time \cite{bb}.
These solutions have the topology of the Euclidean space and so are not useful to establish a connection with Dirac monopoles
when topological charge is of interest.
However, higher dimensional Taub-Bolt solutions are also known \cite{pp,marika}. These are the backgrounds which allow for
non-selfdual Maxwell fields in four dimensions and which possess nontrivial topology, in general.
These higher dimensional configurations have found applications in M-theory when a negative cosmological constant is considered \cite{ac}.

\section{Nuts and Bolts in Gravity and Gauge Theories}
\label{sec:lovelock}

The Einstein--Maxwell solutions we have mentioned until now can be extended to other theories such as Einstein--Born--Infeld
or Lovelock--Maxwell. As a first example of a generalized setting for topological charge and Taub--NUT backgrounds harboring
electromagnetic fields, we refer the interested reader to \cite{dfhq2}. There, it is made manifest that the magnetic flux quantization
is the same as in the Maxwell case. This is expected from the fact that Born--Infeld fields are asymptotically Maxwell.
The values of the electric parameter is indexed by the Chern number as well, but in a way where the nonlinear dynamics influences
the results. Below, we show that this holds also in higher dimensional Einstein or even Lovelock gravity.
From a mathematical point of view we do not need any physical theory for topological results to hold. Notwithstanding,
physically we are interested in reasonable theories where these abstract constructions are realized.
We base our discussion on complex line bundles over K\"ahler manifolds. In four dimensions NUT solutions
with a spherical base space reduce to spherically symmetric spacetimes. In higher dimensions this cannot happen
as no hypersphere allows for a K\"ahler structure. In other words, the static limit of higher dimensional Taub--NUTs
are not spherically symmetric. The work of \cite{ray} provides exact solutions for every dimension and order of Lovelock
theory from which NUT solutions may be constructed. Recently, NUT solutions have been found in theories which are not Lovelock \cite{beyond}.
Our results are not limited to Lovelock--Maxwell theory, however, it is the most general theory of gravity which avoids 
Ostrogradski instabilities by providing, in general, second order equations of motion under any condition \cite{lovelock}. This provides
an ample test bed for the physical realization of our results. Furthermore, any Lovelock theory in four dimensions corresponds
to general relativity.

Before analyzing higher dimensional instantons, we shortly review Lovelock's gravity and other related theories,
where NUT solutions have been found to exist. Afterwards, we emphasize all the characteristics of four dimensional
NUT solutions which will be generalized to higher dimensions.

Five and six dimensional Lovelock theories are alterations of general relativity
in which the Gauss--Bonnet integral contributes to the equations of motion.
In general, the Lovelock theory is given by Euler densities in the Lagrangian.
The Lovelock action in even dimensions $D=2k+2$ is given by
\begin{equation}
 I_L[g]=\int \rd^Dx\sqrt{-g}\left[\sum\limits_{i=0}^{k+1}\alpha_i{\cal L}_i\right],
\end{equation}
where $\alpha_i$ are constants and
\begin{equation}
 {\cal L}_i=\frac{1}{2^i}\delta^{a_{1}a_{2}\dots a_{2i}}_{b_{1}b_{2}\dots b_{2i}}
 R^{\quad~~ b_{1}b_{2}}_{a_{1}a_{2}}\cdots R^{\qquad\quad b_{2i-1}b_{2i}}_{a_{2i-1}a_{2i}}.
\end{equation}
Notice that with increasing dimensionality, higher power curvature terms are included. So, although Lovelock solutions are desirable in higher dimensions, 
they are also increasingly more complex. For the sake of simplicity we consider only a handful of explicit cases, 
which are all described by the following action
\begin{equation}
 I[g,A]=\int \rd^{2k+2}x\sqrt{-g}\left[-2\Lambda + R  +\alpha_2{\cal L}_2+\alpha_3{\cal L}_3 -F_{ab}F^{ab} \right],\label{third}
\end{equation}
where the Maxwell term is included and the Lagrangian densities ${\cal L}_2$ and ${\cal L}_3$ are given explicitly by
\begin{equation}
 {\cal L}_2=R_{abcd}R^{abcd}-4R_{ab}R^{ab}+R^2,
\end{equation}
and
\begin{align}
\begin{split}
 {\cal L}_3=&2R^{abcd}R_{cdef}R^{ef}_{~~ab}+8R^{ab}_{~~cd}R^{ce}_{~~bf}R^{df}_{~~ae}\\
 &+2R^{abcd}R_{cdbe}R^{e}_{~a}+3RR^{abcd}R_{cdab}+24R^{abcd}R_{ca}R_{db}\\
 &+16R^{ab}R_{bc}R^{c}_{~a}-12RR^{ab}R_{ab}+R^3.
\end{split}
\end{align}
In the notation of Eq.(\ref{third}), $\alpha_0=-2\Lambda$ and $\alpha_1=1$.
Instantons of Taub--NUT type are, of course, known for $\alpha_2=\alpha_3=0$ in dimension four.
The next level of complexity is $\alpha_3=0$ with $D=6$ \cite{awad,dehghani1}. Dimension eight follows 
in complexity, a vacuum solution is known for $\alpha_i\neq0$ \cite{hendi}.
Therein, special values of the parameters have also been explored. Special cases of Lovelock
theory include Chern--Simons and Born--Infeld gravity. Frequently explored scenarios are
those where all coupling constants vanish except for the one with the largest index --- sometimes
called pure Lovelock gravity. A third order Lovelock--Maxwell exact solution has recently be obtained in 
\cite{cfq18}.
For any arbitrary even dimension greater than two, NUT geometries are known to exist in Einstein ($\alpha_2=\alpha_3=0$)
and Gauss-Bonnet ($\alpha_3=0$) gravity coupled to Maxwell matter \cite{dehghani1}. 
For a further description of these solutions, we refer to \ref{sec:app}.

In the next section, we compute the Chern index for higher dimensional Taub-Bolt spaces.
We end this section by considering the four dimensional case and highlighting what
point of view is more convenient to interpret the results of their higher dimensional analogues.

The Brill--Carter solution \cite{brill,carter} is the Einstein spacetime which has a possibly nonzero cosmological constant $\Lambda$,
electric and magnetic mass $m,l$ respectively, as well as electromagnetic parameters $Q,p$.
The line element determining the metric is given by
\begin{equation}
 \rd s^2=-f(r)(\rd t+2l\cos\theta \rd\phi)^2+f(r)^{-1}\rd r^2+(r^2+l^2)(\rd\theta^2+\sin^2\theta\rd\phi^2),
\label{brillmetric}
\end{equation}
while the 2-form which locally represents the Maxwell field strength, $F$, is
\begin{equation}
 F=g'(r)\rd r\wedge(\rd t+2l\cos\theta \rd\phi)-g(r)\sin\theta\rd\theta\wedge\rd\phi.
\label{brillfield}
\end{equation}
The functions $f$ and $g$ are given explicitly in equations (\ref{brillf}) and (\ref{brillg}) of \ref{sec:app}.
The geometry can be placed on a manifold with $\mathbb{R}^2\times S^2$ topology but with
physical singularities arising from a topological defect in spacetime. The singularities
can be avoided if one chooses a $\mathbb{R}\times S^3$ underlying manifold topology.
This is at the expense of having closed timelike curves at each point of space. The Brill
solution was constructed as a generalization of Taub spacetime. An anisotropic universe
with spatial homogeneity is given by a three dimensional hypersphere. It is a very special type
of Bianchi IX model which possesses an additional isometry. This symmetry is biaxial and
reserves the roundness of the two-sphere in the Hopf fibration of spatial slices. This
sector of the Taub--NUT space generalizes the interior of the Schwarzschild solution.
When it is extended through a null hypersurface, the Hopf fibered direction becomes timelike.
Instantons which arise from Taub--NUT geometry are analytic continuations which have adopted
the $\mathbb{R}\times S^3$ topology. Not until very recently \cite{Lthd} this was the only way in which
consistent thermodynamics was produced for NUT spacetimes. We comment more on the thermodynamics
of NUT solutions below.

In this work we adopt a global structure of the spacetime manifold which has no observable topological defects.
In this Taub--NUT geometry a compact direction is Hopf-fibered over a sphere.
This base space happens to be a K\"ahler manifold. The K\"ahler structure of the round metric
is $\omega=\sin\theta\rd\theta\wedge\rd\phi$. By Poincar\'e's lemma the form is locally represented
by the exterior derivative of a one-form potential, in this case $B=-\cos\theta\rd\phi$.
In Eq.(\ref{brillmetric}), we have written the metric in Boyer-Lindquist coordinates.
The nondiagonal term in the asymptotic metric is determined by
\begin{equation}
 2l\rd t B=2l\cos\theta\rd t\rd\phi.
\end{equation}
which suggests an angular momentum interpretation of the NUT parameter $l$. Notice how the underlying K\"ahler
structure determines the form of the spacetime geometry and how it couples to the NUT parameter.
This carries over to the Maxwell sector when aligned fields are considered.
The gauge potential can be chosen as $A=g(r)(dt+2lB)$ rendering equation (\ref{brillfield}), as $F=\rd A$.
At the asymptotic boundary ($r=\infty$) The field strength is determined by the K\"ahler two-form $\omega$
\begin{equation}
 F_{\infty}=-g(\infty)\omega. \label{Fomega}
\end{equation}
The K\"ahler submanifold is a sphere and so the field strength at the boundary corresponds to a Dirac monopole.
Let us consider a circular bundle over spacetime and in which the connection of the bundle space is 
locally represented by the U(1) gauge potential. Then there is a subbundle where the curvature of the gauge potential
is given by (\ref{Fomega}). This is the topological structure of a Dirac monopole whose topological charge is
$2g(\infty)$. However, the Maxwell field on all spacetime has vanishing topological charge because of the topology we chose.
The complementary topology allows for a nonvanishing topological index but also introduces conical singularities.
However, it has been found that analytical continuations of the spacetime do not possess any of these obstructions \cite{dfhq}.

A first step towards finding Euclidean solutions of this type is a Wick rotation $t\to \ri\tau$, immediately afterwards
it is found necessary to rotate the NUT parameter, say $l\to \ri n$. This final change guarantees a real Riemannian metric.
The line element now takes the form
\begin{equation}
 \rd s^2=Y(r)(\rd \tau+2n\cos\theta \rd\phi)^2+Y(r)^{-1}\rd r^2+(r^2-n^2)(\rd\theta^2+\sin^2\theta\rd\phi^2).\label{wickmetric}
\end{equation}
The exact function $Y$ depends on the gravitational theory where the geometry is realized. The same is true for the gauge potential
which depends on the specified electrodynamics. However, if the background is (\ref{wickmetric}), even without a specified gravity theory,
then the Maxwell equations are completely determined, independently of the function $Y$, for a gauge potential of the form
\begin{equation}
 A=h(r)(\rd\tau+2nB).\label{wickgauge}
\end{equation}
The Lorentzian metric (\ref{brillmetric}) has null hypersurfaces when the metric function degenerates.
Its Euclidean version (\ref{wickmetric}), instead, has degenerate submanifolds which correspond to the roots of $Y$.
Notice that these submanifolds are the fixed point set of the circular action of $\partial_{\tau}$.
If this set is a point, it is called a \emph{nut}, which 
happens when $r=n$. The metric completely degenerates, but this point is regular. So, the global topology is
$\mathbb{R}^4$ and the gauge potential can be defined globally with one open cover. It also follows that the only fiber bundle which
can be constructed over it is the trivial one. Complementary when the metric degenerates at $r=r_+>n$ it does 
so into a two-dimensional sphere. This fixed point set of codimension two
is called a \emph{bolt}. The global topology coincides with that of $\mathbb{CP}^2\char`\\\{*\}$. Notice that the space is the fibration
of a plane over the bolt and, therefore, it has the homotopy type of the bolt.
In the presence of a NUT parameter gravitational instantons with a bolt are called Taub-Bolt.

In Euclidean quantum gravity,  instantons are important because they correspond to saddle points of the gravitational path integral.
The integral has fixed boundary conditions with a given topology and geometry. Varying infilling topologies are limited by selecting only
geometries with a desired symmetry. If we choose a spherical boundary with ${\mathfrak{su}}(2)\oplus {\mathfrak{u}}(1)$ isometry algebra,
then Taub--NUT and Taub-Bolt solutions are saddle points of the Einstein action. This analysis also applies for the field content.
For these metrics the ${\mathfrak{su}}(2)$ isometry corresponds to a homogeneity algebra and the additional ${\mathfrak{u}}(1)$ isometry
guarantees the roundness of the K\"ahler base. These features are detailed here because they carry over to higher dimensions.
An additional consideration which comes from the Maxwell sector is that
the electric charge must be Wick rotated as well, the reason is as before, to guarantee a real valued field.
As a final comment we point out that saddle points of any path integral require the manifolds to be regular on every point.
In other words, infilling spacetimes with curvature singularities are inadmissible. In fact, avoidance of conical singularities
along the Euclidean time direction at the degenerate submanifolds generically defines the ensemble's temperature.
These regularity conditions are crucial for physical applications.

\section{A Higher Dimensional Reprise}

Inhomogeneous Einstein metrics built over complex line bundles are our main interest \cite{pp}. The geometries under consideration have the generic form
\begin{equation}
\rd s^2 = \texttt{} Y(r)(\rd \tau+2nB)^2+Y(r)^{-1}\rd r^2+(r^2-n^2)\rd\Sigma^2, \label{inhommetric}
\end{equation}
where $\rd\Sigma^2$ is a K\"ahler metric line element with corresponding K\"ahler structure given by $\omega=\rd B$.
The complex dimension of the K\"ahler space is $k$ so that the total real dimension is $D=2k+2$.
We specialize to the case where the base space is compact. The complex line fiber has a periodic coordinate
identified with Euclidean time. The bundle space represents a Euclidean spacetime with asymptotic boundary
that has the topology of a Hopf bundle over the K\"ahler base. By definition the second cohomology
group of the base is nontrivial, the Hopf bundle is the one which has unit Chern index. Any isometry of the base space
is also an isometry of the total space. In addition  to those possible symmetries, the metric (\ref{inhommetric}) has a
circular isometry coming from the Euclidean time; its periodicity is fixed by boundary conditions.
These conditions must be consistent with the definition of the ensemble temperature, as mentioned in the previous section.
Notice that only when the boundary is spherical, it contains Euclidean space as an infilling topology.
The radial direction of the complex line fiber, $r$, determines a foliation of the total space. 
The warping factor of the K\"ahler sector indicates that when
$r=n$ the K\"ahler submanifold degenerates. If the circular direction does so too for this values of $r$, then it marks a singularity
of the manifold. This is unless the Hopf bundle continuously degenerates into a point. This only happens when the boundary is spherical, as spheres
can continuously crush into a point.
In other words, regular manifolds with this metric which carry a single nut (at $r=n$) exist only when the boundary is spherical.
These solutions are the higher dimensional Kaluza--Klein monopoles mentioned earlier. The existence of these solutions implies that the
base manifold must be a complex projective space.

The previous discussion has fixed the class of metrics we investigate in this work.
The base space is in fact an Einstein--K\"ahler manifold and the K\"ahler one-form potential
is denoted by $B={\cal A}_k$. 
Both the potential and the line element $\rd\Sigma_k^2$ can be defined recursively \cite{hoxha}. 
These definitions exploit the fact that every space $\mathbb{CP}^k$ contains an extrinsically flat submanifold $\mathbb{CP}^{k-1}$.
First, let us define
\begin{equation}
 {\cal A}_1=2\sin^2\psi_1 \rd\phi_1=(1-\cos\theta_1) \rd\phi_1.
\end{equation}
The coordinate $\phi_1$ varies from 0 to 2$\pi$ while $\psi_1$ does so from 0 to $\pi$/2 and $\theta_1=2\psi_1$.
The coordinates ($\psi_1,\phi_1$) best accommodate the recursive definitions mentioned above. The pair
($\theta_1,\phi_1$) are spherical coordinates and are presented for comparative reasons.
Since $\mathbb{CP}^1$ coincides with $\mathbb{S}^2$, we have that
\begin{equation}
 \rd\Sigma_1^2=4(\rd\psi_1^2+\sin^2\psi_1\cos^2\psi_1\rd\phi_1^2)=\rd\theta_1^2+\sin^2\theta_1\rd\phi_1^2.
\end{equation}
The four dimensional case is readily seen to be the Taub--NUT space --- for the appropriate function $f(r)$.
Since the potential ${\cal A}_1$ is not well defined everywhere on the sphere, for a global definition over the background
space we should use 
\begin{equation}
 {\cal A}'_1=2\cos^2\psi_1 \rd\phi_1=(1+\cos\theta_1) \rd\phi_1.
\end{equation}
Thus, the recursive formulae read
\begin{align}
 {\cal A}_k&=(k+1)\sin^2\psi_k\left(\rd\phi_k+\frac{1}{k}{\cal A}_{k-1}\right),\\
 \rd\Sigma_k^2&=2(k+1)\left[\rd\psi_k^2+\sin^2\psi_k\cos^2\psi_k\left(\rd\phi_k+\frac{1}{k}{\cal A}_{k-1}\right)^2 +\frac{1}{2k}\sin^2\psi_k\rd\Sigma_{k-1}^2\right].
\end{align}
Now that the base space has been fixed, the Euclidean spacetime metric has an isometry algebra ${\mathfrak{su}}(k+1)\oplus{\mathfrak{u}}(1)$.
The topology fills in the interior of a spherical boundary $S^{2k+1}$. Depending on the fixed point set of the circular action of Euclidean time,
it will either be Euclidean space $\mathbb{R}^{2k+2}$ or $\mathbb{CP}^k\char`\\\{*\}$ --- A Taub--NUT or a Taub-Bolt space.
As for the matter content, Eq.(\ref{wickgauge}) now takes the form
\begin{equation}
 A=h(r)(\rd\tau+2n{\cal A}_k), \label{gauge}
\end{equation}
for each specified dimension.
Recall that $h$ is a function which does not vanish asymptotically, i.e., $h(\infty)=v$.
For Maxwell theory the function is given explicitly in equation (\ref{ldh}) of \ref{sec:app}.
In dimensions higher than four, it is not
obvious that this quantity is of magnetic nature. This contrasts with its electric counterpart which is perfectly well defined
in any dimension. Still, we consider $v$ in higher dimensions as a magnetic quantity and justify this assumption below.
This initial setup paints the picture for carrying out the thermodynamics of these configurations.
A highly nontrivial task which has has advanced from works such as \cite{emparan, kraus, taylor} and many others.
The thermodynamics of instantons with magnetic mass has been considered in Einstein--Maxwell theory \cite{dehghani2}
and in Gauss--Bonnet--Maxwell \cite{dehghani1}. Higher curvature theories such as Gauss--Bonnet or Lovelock, in general,
allow for interesting physical features such as negative entropy in the presence of a cosmological constant \cite{cvetic}.
Extended thermodynamic scenarios have also been formulated
for these type of instantons \cite{cvj,vol}, where negative entropy has been found for certain regimes of AdS--Taub--NUT.

The higher dimensional Euclidean Taub--NUT solutions or Kaluza--Klein monopoles above admit generalizations from vacuum solutions
to electrovacuum. It has been emphasized that the metric is regular at the nut and that the field must be regular there as well.
The regularity condition is then $h(n)=0$ so that the gauge potential vanish at the nut as the Euclidean time direction degenerates.
The potential difference between the nut and the boundary is given then by the value of the function $h$ at infinity.
This yields a physical interpretation for the integration constant $v$. Examining the asymptotic behavior of the field strength
shows $q$ to be the electric charge (up to a rescaling). However, regularity does not allow these integration constants to be independent.
Finally, we comment that these backgrounds are all homotopic to a point and, thus, no topological charge is present. 

Taub-Bolt instantons in any dimension have the homotopy type of their respective bolt, which are K\"ahler and so
have nontrivial second order cohomology. This signals topological charges for U(1) gauge fields. For spherical
boundary topology the bolts are complex projective spaces. The circular bundles which represents the electromagnetic fields
all possesses a subbundle which is isomorphic to the Dirac monopole structure. It is unique in the sense that all spaces $\mathbb{CP}^k$
have one and only one totally geodesic $\mathbb{CP}^1$ submanifold. The magnetic field at infinity has a nontrivial flux through this special
submanifold. Restricted to this sphere at infinity the field corresponds to a Dirac monopole. The circular subbundle over this sphere 
has a Chern number
\begin{equation}
 -\frac{1}{2\pi}\int\limits_{\mathbb{CP}^1_{\infty}}F_k=(k+1)2nv\equiv 2p.\label{ChernDirac}
\end{equation}
This provides $v$ with a magnetic interpretation for higher dimensional Taub-Bolt instantons.
The factor in front of $v$ depends on the dimension. The total geodesic subspace is determined by
fixing $\psi_k=\pi/2$ for all $k>1$. This topological index is related to an underlying Brouwer degree
from the sphere's covering intersection to the U(1) gauge group. It is a winding number $w$
\begin{equation}
 2p=w=1,2,3\dots
\end{equation}

We now proceed to calculate the Chern number (cf. \cite{naber2}) associated with the self gravitating Euclidean Maxwell field.
Given that the gauge group is unitary of degree one, then its only nonvanishing Chern number is
\begin{equation}
 c_1^{k+1}[M]=\int\limits_{M}c_1^{\wedge(k+1)},
\end{equation}
where $c_1=F/2\pi$ and $F$ is the field strength of the potential (\ref{gauge}). 
The configurations are completely classified by this topological index \cite{naber2}.
We write this number as a ${\mathfrak{u}}(1)$ Chern--Simons invariant on the boundary taking the usual induced orientation
\begin{equation}
 cs[\partial M]=\frac{1}{(2\pi)^{k+1}}\int\limits_{S^{2k+1}_{\infty}}A\wedge F^{\wedge k}.
\end{equation}
The periodicity of $\tau$ is well known to be $\beta=4\pi n(k+1)$, so as to avoid conical singularities \cite{marika,ac}.
Then,  we make a first simplification
\begin{equation}
 cs\left[S^{2k+1}_{\infty}\right]=\frac{v\beta(2nv)^k}{(2\pi)^{k+1}}\int\limits_{\mathbb{CP}^k}{\cal F}_k^{\wedge k},
\end{equation}
where ${\cal F}_k=\rd{\cal A}_k$ has the recursive form
\begin{equation}
 {\cal F}_k=(k+1)\left[\rd(\sin^2\psi_k)\wedge\left(\rd\phi_k+\frac{1}{k}{\cal A}_{k-1}\right)+\sin^2\psi_k\frac{1}{k}{\cal F}_{k-1}\right].
\end{equation}
The wedge product considerably simplifies the expression ${\cal F}_k^{\wedge k}$. In the above equation, the sum to the wedge power of $k$
has only one surviving term of the binomial expansion and so it yields
\begin{equation}
 {\cal F}_k^{\wedge k}=(k+1)^kk(\sin^2\psi_k)^{k-1}\rd(\sin^2\psi_k)\wedge\rd\phi_k\wedge\left(\frac{{\cal F}_{k-1}^{\wedge k-1}}{k^{k-1}}\right).\label{iter}
\end{equation}
By successive iterations of equation(\ref{iter}), we have
\begin{equation}
 \int\limits_{\mathbb{CP}^k}{\cal F}_k^{\wedge k}=(k+1)^{k}(2\pi)^{k},
\end{equation}
giving us the  index
\begin{equation}
 c_1^{k+1}=[2nv(k+1)]^{k+1}=w^{k+1}.
\end{equation}
In higher dimensions there are many magnetic fluxes, but the one which can be read off from (\ref{ChernDirac}) is special as it can only
take certain discrete values, like a Dirac monopole \cite{naber3}. It also closely resembles the topological charge in magnetic
flux tubes (cf. \cite{bais}). The magnetic charge is called topological as any value of $p$ determines the topology of
the bundle space.

The Taub-Bolt spaces are defined to be regular manifolds so various conditions have to hold. The periodicity of Euclidean time
is forced by the boundary condition to be $\beta=4\pi n(k+1)$. However, regularity of the bolt, say $r=r_b$, requires it to be $\beta=4\pi/Y'(r_b)$.
The bolt's area is restricted by these conditions. When the background harbors a self-gravitating aligned electromagnetic field, 
then the additional condition $h(r_b)=0$ contributes to the determination of allowed bolt sizes. When the field is of Maxwell type,
the electric charge is fixed by this condition and by Eq.(\ref{ldh}); in other words
\begin{equation}
 q_w=(-1)^{k+1}\frac{wn^{2k-1}}{2r_b}\frac{2k-1}{k+1}~_2{\rm F}_1(-1/2,-k;1/2; r_b^2/n^2). \label{qcondition}
\end{equation}
The proportionality between the electric and magnetic parameters varies according to the dimension, the NUT parameter and the size of
the bolt. To sum up: the electric charge is indexed by the topological charge of the Maxwell field $w/2$.
However, this result is not topological its origin is Euclidean quantum gravity.

\section{Discussion}

We have shown that for arbitrary even dimension Taub--NUT dyons are analogues of the Dirac monopole.
Their self-gravity produces backgrounds which have nontrivial topology. They are classified by a single
topological index which corresponds to a gauge field parameter. This integration constant is the magnetic charge
in four dimensions. In dimensions higher  than four, the bundle structure of the entire Euclidean spacetime
is rooted in a Dirac monopole fibration. This result crucially depends on the nonvanishing of the NUT parameter.
The Chern number corresponds to a power of the root monopole's topological
charge. In reference \cite{dfhq}, it came as a surprise that electric charge was quantized as no topological reason
exists for this condition. However, the quantization condition is a consequence of Euclidean quantum gravity
in the semiclassical regime. Here, we have generalized this result to even dimensions higher than four.
This result is succinctly written in equation (\ref{qcondition}). This equation relies on the topological
indexing of the integration constant $v$, which applies in general. Nonetheless, the semiclassical electric charge
quantization requires further specification, such as the geometry.

Our results parallel some notions of the Dirac monopole and so we can compare them to the Dirac quantization condition.
The condition relates an electric current, which interacts with the magnetic gauge field.
Taub-NUT spaces are spin$^{\mathbb{C}}$ manifolds thus charged spinor fields can be defined
on them. So one expects a Dirac conditioning of the type
\begin{equation}
2pq_F=\hbar
\end{equation}
where $p$ is the magnetic flux per solid angle of the Maxwell field and $q_F$ is the electric charge
of the spinor, the coupling constant in the action. However, the result we have focused on is the indexation of the electric charge
which comes from the field itself. The dyonic configuration has a discrete magnetic sector because of topological reasons.
The electric sector is affected by this through the regularity condition imposed by the path integral approach
to quantum gravity. So both sectors of the dyon are ultimately indexed by the same winding number.

In a setting where the thermodynamics of these dyons are expected to hold, some concerns arise because of the discrete jumps the magnetic flux exhibits. 
However, the magnetic flux is in correspondence with the electric potential at infinity. In the fixed potential ensemble, 
the electric charge varies with respect to temperature, for example, but manifests no discrete jumps because its index is fixed. Moreover,
in this ensemble Taub--NUT spaces transition into Taub-Bolt phases \cite{cvj}. 
Since Taub-Bolt spaces require the magnetic flux to be quantized, then any Taub--NUT space which transitions
into them in the fixed potential ensemble must have a flux per solid angle which is a half integer as well.
The very definition of the NUT ($r_+=n$) and Bolt ($r_+>n$) phases entails that during this transition
there is a discontinuity in the electric charge. Recall that the electric charge is indexed by the relation $h(r_+)=0$.
This discontinuity is characteristic of first order phase transitions.

To sum up, higher dimensional Kaluza--Klein monopoles are a topologically trivial Taub--NUT space,
which is related to Dirac monopoles through dimensional reduction. The complementary spaces to these higher dimensional
solutions are Taub-Bolt spaces. The novel aspect of our work is that it connects these latter instantons with the Dirac
monopole through algebraic topology. In the process, a unique magnetic flux is singled out in the Taub-Bolts, because it
exhibits topological quantization. This indirectly causes a quantization of electric charge.
The Taub-Bolts are both topologically and energetically the stable phase of NUT/Bolt systems, so they represent
the long-lived configuration.

\section*{Acknowledgments}

We would like to thank Crist\'obal Corral for useful comments and discussions. 
DFA is supported by CONACyT under Grant No. 404449. 
This work was partially supported  by UNAM-DGAPA-PAPIIT, Grant No. 111617, 
and by the Ministry of Education and Science of RK, Grant No. 
BR05236322 and AP05133630.

\appendix

\section{Explicit Solutions}
\label{sec:app}

The magnetic flux quantization and topological invariant associated with the higher dimensional
self-gravitating electromagnetic fields investigated in this work do not depend on the dynamics or geometry
of the background. For the sake of completeness, we present here some  explicit solutions,
which arose during the elaboration of the present work. 

The explicit form of the functions $f$ and $g$ given in the equations (\ref{brillmetric}) and (\ref{brillfield}) is
\begin{subequations}
 \begin{align}
 f(r)&=\frac{r^2-2mr-l^2+Q^2+p^2}{r^2+l^2}\label{brillf}\\
 g(r)&=\frac{Qr+(p/2l)(r^2-l^2)}{r^2+l^2}.\label{brillg}
 \end{align}
\end{subequations}
A Wick rotation relates the previous functions with $Y$ and $h$, as given in (\ref{wickmetric}) and (\ref{wickgauge}), which then become
\begin{subequations}
 \begin{align}
 Y(r)&=\frac{r^2-2mr+n^2-q^2+p^2}{r^2-n^2}\label{brilly}\\
 h(r)&=\frac{-qr+v(r^2+n^2)}{r^2-n^2}.\label{brillh}
 \end{align}
\end{subequations}
Here, the electric charge $Q$ has been Wick rotated into $\ri q$ and we have set $v=p/2n$.

The Lovelock solutions described in Sec. \ref{sec:lovelock} have a geometry given by equation (\ref{wickmetric}).
We quote $Y(r)$ and $h(r)$ in (\ref{gauge}) for the four dimensional case
\begin{subequations}
 \begin{align}
 Y(r)&=\frac{r^2-2mr+n^2-3^{-1}\Lambda(r^4-6n^2r^2-3n^4)-q^2+p^2}{r^2-n^2},\label{l4Y}\\
 h(r)&=\frac{-qr+v(r^2+n^2)}{r^2-n^2}.\label{l4h}
 \end{align}
\end{subequations}
The six dimensional case is given in \cite{dehghani1}
\begin{subequations}
 \begin{align}
 Y(r)&=\frac{(r^2-n^2)^2}{12\alpha_2(r^2+n^2)}\left(1+\frac{4\alpha_2}{r^2-n^2}-\sqrt{B(r)+C(r)}\right),\label{l6Y}\\
 h(r)&=\frac{-qr+v(r^4-6n^2r^2-3n^4)}{(r^2-n^2)^2}.\label{l6h}
 \end{align}
\end{subequations}
with the auxiliary functions defined by
 \begin{align}
  B(r)&=1+\frac{16\alpha_2n^2(r^4+6n^2r^2+3n^4)+12\alpha_2mr(r^2+n^2)}{(r^2-n^2)^4}+\notag\\
 &\quad+\frac{12\alpha_2\Lambda(r^2+n^2)(r^6-5n^2r^4+15n^4r^2+5n^6)}{5(r^2-n^2)^4}+\notag\\
 &\quad+\frac{3\alpha_2(r^2+n^2)(4n^3(3r^2-n^2)q^2-128n^5r^3qv)}{n^3(r^2-n^2)^6}+\\
 &\quad+\frac{96\alpha_2n^5(r^2+n^2)(r^6+15n^2r^4-9n^4r^2+9n^6)}{n^3(r^2-n^2)^6},\notag\\
 C(r)&=\frac{\alpha_2^2[-16(r^4+6n^2r^2+n^4)+3r(r^2+n^2)q^2/(\alpha_2n^3)]}{(r^2-n^2)^4}.\notag.
\end{align}
The purely gravitational eight-dimensional analogue leads to a cumbersome metric function $Y(r)$, which has been partially 
presented in \cite{hendi}  for a particular choice of  the coupling constants.  
Moreover, we have generalized this result to incorporate
an aligned Maxwell field \cite{cfq18} such as in (\ref{gauge}) with
\begin{equation}
 h(r)=\frac{-qr+v(r^6-5n^2r^4+15n^4r^2+5n^6)}{(r^2-n^2)^3}.\label{l8h}
\end{equation}

In dimension $2k+2$, for the geometry given by (\ref{wickmetric}) these gauge fields are given by \cite{dehghani1}
\begin{equation}
 h(r)=\frac{1}{(r^2-n^2)^k}\left[-qr+v(-1)^{k+1}(2k-1)n^{2k}~_2{\rm F}_1(-1/2,-k;1/2; r^2/n^2)\right].\label{ldh}
\end{equation}
Notice that the values $h(\infty)=v$ and $h(r_+)=0$ lead to a value of $q$ which is proportional to $p$.

\end{document}